\documentclass[a4paper,12pt]{article}
\usepackage[a4paper,width=150mm,top=25mm,bottom=25mm,bindingoffset=6mm]{geometry}
\usepackage[english]{babel}
\usepackage[pdftex]{graphicx}
\usepackage{placeins}
\usepackage{amsmath,amsthm,amsfonts,amssymb,mathtools,bm,latexsym,stmaryrd}
\usepackage{rotating}
\usepackage{tabularx}
\usepackage[table,dvipsnames]{xcolor}
\usepackage{multicol,multirow}
\usepackage{relsize}
\usepackage[center]{subfigure}
\usepackage{caption}
\usepackage{slashed}
\usepackage{lineno}
\usepackage{cite,mcite}
\usepackage[title]{appendix}
\usepackage{url}
\usepackage[bookmarks=false,colorlinks]{hyperref}
\hypersetup{urlcolor=NavyBlue,citecolor=NavyBlue}
\usepackage{braket}

\newcommand{\ketbra}[2]{{|{#1}\rangle\langle{#2}|}}

\def\Y{\mathcal{Y}}
\def\Xp{{X^\prime}}
\def\Yp{{Y^\prime}}

\def\ph{\text{ph}}

\def\erf{\text{erf}}

\def\rhobar{{\overline{\rho}}}

\def\phibar{{\overline{\varphi}}}
\def\thetabar{{\overline{\theta}}}

\begin{document}

\title{\bf\Large Imperfect preparation and Trojan attack on the phase modulator in the decoy-state BB84 protocol}
\author{\sc Aleksei Reutov}

\date{}%\today}
\maketitle
\thispagestyle{empty}

{\centering \sl\small \noindent
QRate, Moscow, Russia\\ Moscow Center for Advanced Studies, Moscow, Russia\par}

\vskip4cm
\begin{abstract}
    Quantum key distribution (QKD) provides a theoretically secure method for cryptographic key exchange by leveraging quantum mechanics, but practical implementations face vulnerabilities such as Trojan horse attack on phase modulators. This work analyzes the security of QKD systems under such attacks, considering both ideal and imperfect state preparation scenarios. The Trojan attack model is generalized to arbitrary states of probing pulses and conservative bounds of information leakage through side-channel of special form are introduced. The quantum coin imbalance, a critical security parameter, remains low (on the order of $10^{-7}$ for ideal state preparation and $10^{-5}$ for imperfect preparation) with this new approach and presence additional hardware passive countermeasures. Numerical simulations confirm nonzero secure key rate at distances over 100 km through optical fiber channel. 
    % The quantum key distribution (QKD) allows two remote users to share a common information-theoretic secure secret key. In order to guarantee the security of a practical QKD implementation, the physical system has to be fully characterized and all deviations from the ideal protocol due to various imperfections of realistic devices have to be taken into account in the security proof. In this report, we study the security of the efficient decoy-state BB84 QKD protocol in the presence of source flaws, caused by imperfect intensity and polarization modulation. We investigate 
    % %the 
    % non-Poissonian photon-number statistics due to coherent-state intensity fluctuations and the basis-dependence of the source due to non-ideal polarization state preparation. The analysis is supported by experimental characterization of intensity and phase distributions.
    % \begin{description}®r
    % \item[Keywords:] quantum communication, quantum key distribution, QKD, information leakage, security analysis, source flaws.
    % \end{description}
\end{abstract}

\pagebreak
\section*{Introduction}

Quantum key distribution (QKD) represents a main area of modern quantum cryptography, suggesting a theoretically secure method \cite{BB84} for exchanging cryptographic keys between distant parties. By leveraging the principles of quantum mechanics, an ideal implementation of QKD ensures that any eavesdropping attempt on the key exchange process is detectable, thereby providing a level of security unattainable by classical cryptographic methods. However, the practical implementation of QKD systems is not without challenges \cite{Xu20}. Real-world devices often deviate from assumptions of idealized theoretical QKD schemes, introducing vulnerabilities that can be exploited by sophisticated attacks. Among these, the Trojan horse attack \cite{Vakhitov01} on phase modulators stands out as a powerful threat, capable of compromising the security of QKD systems without leaving detectable influence on observables measured during QKD protocol.

% This article addresses the critical issue of Trojan horse attacks in QKD systems, focusing on their impact on the security of the key distribution process. We begin by examining the state preparation process, which is fundamental to the generation of secure quantum states. Any deviation from the ideal preparation of polarization states, such as imperfections in the azimuth and phase angles, can lead to non-orthogonal states and introduce vulnerabilities. These imperfections are modeled using a Gaussian approximation, providing a realistic framework for analyzing the security of practical QKD systems.

This article addresses the BB84 decoy-state protocol \cite{Trushechkin17, Trushechkin21} and the specifics of Trojan-horse attacks on phase modulators (PMs). Unlike some previous studies \cite{Lucamarini15, Wang18, Tamaki16} that assume coherent states for the probing Trojan pulses, this article is appealed to a more general scenario where the eavesdropper can prepare arbitrary states for phase-modulator's probing. Such generalization allows to derive conservative bounds on the information leakage caused by Trojan attacks and Alice's side channels. Side channel of specific and low-dimensional form is proposed as upper bound for any Trojan probing by arbitrary (pure or mixed) state. This result is formalized in three Statements, one for asymptotic case and two other for finite-key effects with statistical corrections based on Chernoff bounds. The possible imperfect state preparations are also considered and the imperfect choice of bits and bases are modeled using a Gaussian approximation close to behavior of quantum-state sources in practical QKD systems.

Presented in this article analysis reveals that the quantum coin imbalance, a critical parameter affecting phase error, remains weakly varying in the presence of Trojan attack and without it. There are demonstrated for realistic case of QKD implementation with isolators and spectral filters that the quantum coin imbalance values are on the order of $10^{-7}$ for ideal state preparation and $10^{-5}$ for imperfect preparation. These findings are validated through numerical simulations of the secret key rate, which show that our conservative approach yields low changes in key rate and maximum transmission distance. Furthermore, the performance of the protocol under realistic conditions of practical schemes is evaluated and it is demonstrated that secure key distribution is achievable over distances higher than 100 km.

% The results presented in this article not only advance our understanding of the vulnerabilities associated with Trojan horse attacks but also provide practical tools for enhancing the security of QKD systems. By replacing arbitrary Trojan emissions with a conservative side channel model, we establish a robust framework for evaluating the security of QKD protocols in the presence of realistic imperfections. This work contributes to the ongoing effort to bridge the gap between theoretical security proofs and practical implementations, paving the way for the widespread adoption of quantum-secured communication networks.

In the following sections, an exposition of the state preparation process (Appendices \ref{app:imp_prep} and \ref{app:gauss} contain more details about imperfect state preparation and Gaussian approximation for it) is provided, the theoretical framework for Trojan attack on PM is analyzed and the numerical results support article`s conclusions. This findings underscore the importance of rigorous security analysis in the design and implementation of practical QKD systems, ensuring their resilience against both known and emerging threats.

\section{State preparation}
\label{sec:phase}

\begin{figure}[t!]\centering
	\includegraphics[width=0.45\textwidth]{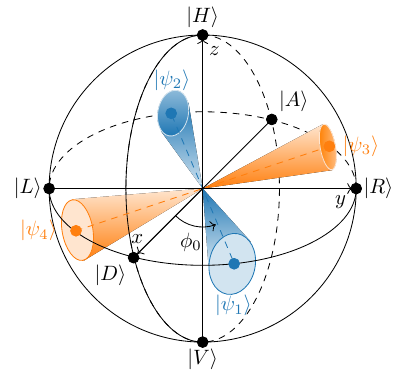}
	\caption{Polarization states on the Bloch sphere transmitted by Alice, reprinted from \protect\cite{Reutov23}. The blue and orange dots mark the perfectly prepared \eqref{eq:psi_ideal} states and the \eqref{eq:psi_real} states are schematically depicted as orange and blue areas on the sphere surface.}
	\label{fig:Poincare_sphere}
\end{figure}

The variation of BB84 protocol with decoy-state technique \cite{Trushechkin17, Trushechkin21} is used in this paper. BB84 \cite{BB84}, introduced by Bennett and Brassard in 1984, is a quantum key distribution (QKD) method that enables two distant parties, Alice and Bob, to securely generate a shared secret key using the principles of quantum mechanics. Alice sends qubits to Bob encoded in one of two non-orthogonal bases and Bob measures them in a randomly chosen basis, after which they publicly compare bases to discard mismatched results. The final key is distilled through error correction and privacy amplification providing information-theoretical security.

The decoy-state technique \cite{Hwang03,Lo05, Wang05, Ma05} is a method to enhance security and key generation rates in practical implementations using weak coherent pulses, which are susceptible to photon-number-splitting (PNS) attacks. By randomly adding decoy pulses of varying intensities to signal pulses, Alice and Bob can estimate the contribution of single-photon and multi-photon measurements. It allows two distant parties to detect and handle with eavesdropper's (Eve's) PNS attacks. This technique significantly improves the robustness and performance of QKD systems under realistic conditions.

In our implementation of decoy-state BB84 protocol \cite{Duplinsky17} we use polarization encoding of prepared states. Any arbitrary pure polarization state can be described by spherical coordinates $\varphi\in[0,2\pi)$ and $\theta\in[0,\pi]$ on the Bloch sphere,
\begin{equation}
	\ket{\psi(\varphi,\theta)} = \cos\bigg(\frac{\theta}{2}\bigg) \ket{H} + e^{i\varphi} \sin\bigg(\frac{\theta}{2}\bigg) \ket{V} \,,
\end{equation}
where $\ket{H}$ and $\ket{V}$ denote the horizontal and vertical polarization states respectively. The azimuth angle is set to $\theta=\pi/2$ in the protocol and the relative phase, which determines the basis and the bit value, is adjusted to $\phi_i\in\{0,\pi,\pi/2,3\pi/2\}$. Thus, the azimuth angle of the $i\text{th}$ state is $\varphi_i=\phi_0+\phi_i$.

States in our practical scheme are prepared and measured (with assumption of ideal phase modulators) in elliptical polarization bases $\Xp:\{\ket{\psi_1},\ket{\psi_2}\}$ and $\Yp:\{\ket {\psi_3},\ket{\psi_4}\}$ obtained by rotating the basis vectors $X:\{\ket{D},\ket{A}\}$ and $Y:\{\ket{R },\ket{L}\}$ around the $z$-axis by $\phi_0$ (see Fig.~\ref{fig:Poincare_sphere}):
\begin{equation}
	\ket{\psi_{1,2}^\text{perfect}} = \frac{1}{\sqrt2} \big( \ket{H} \pm e^{i\phi_0} \ket{V} \big) \,, \quad
    \ket{\psi_{3,4}^\text{perfect}} = \frac{1}{\sqrt2} \big( \ket{H} \pm i e^{i\phi_0} \ket{V} \big) \,.
	\label{eq:psi_ideal}
\end{equation}
The density matrices of these states can be expressed as:
\begin{equation}
	\rho_i = \ketbra{\psi_i^\text{perfect}}{\psi_i^\text{ perfect}}\,.
	\label{eq:rho_ideal}
\end{equation}
Bob will perform measurements in the bases $\Xp$ and $\Yp$ randomly applying one of the positive operator-valued measures $\{\ketbra{\psi_i^\text{perfect}}{\psi_i^\text{ perfect}}\}$.

However, phase modulators a not ideal in practical schemes (e.g., due to imperfect voltage control or mechanical inaccuracy of connection between the optical components \cite{Reutov23}). Thus, there is a deviation from $\theta=\pi/2$ on the Bloch sphere, as well as deviations of $\varphi_i$ from the ideal $\varphi_i\in\{0,\pi,\pi/2,3\pi/2\}$. As a result of all these imperfections, the ideal states \eqref{eq:psi_ideal} change and can be written as:
\begin{equation}
	\ket{\psi_i} = \ket{\psi(\varphi_i, \theta)} \,,
	\label{eq:psi_real}
\end{equation}
where $\varphi_i$ and $\theta$ are random variables with some probability distributions. Note that, these states are no longer mutually orthogonal in the corresponding basis and do not lie in the $xy$ plane of the Bloch sphere. In Appendix \ref{app:imp_prep} and \ref{app:gauss} the case of imperfect preparation is considered in more details and the density matrices $\rhobar_1$, $\rhobar_2$, $\rhobar_3$ and $\rhobar_4$ are determined for the four polarization states of the protocol instead of $\ket{\psi_i^\text{perfect}}$.

\section{Trojan attack on phase modulator}
Trojan probing \cite{Vakhitov01} of phase modulators in QKD schemes refers to an attack strategy where Eve actively interrogates the phase modulation components to extract sensitive information about the quantum states being transmitted. Eve can be remaining undetected by conventional security measures during this attack. Possible robust countermeasures includes real-time monitoring, anomaly detection, passive isolation of QKD devices by outer probing emission and modifications of theoretical protocol.

A number of papers \cite{Lucamarini15, Wang18, Tamaki16} consider the Trojan attack under the assumption that the probe pulses are coherent states $\ket{\alpha}$ (the polarization dependencies are omitted for simplicity):
\begin{equation}
    \ket{\alpha} = e^{-|\frac{\alpha^2|}{2}}\sum_{n=0}^{\infty} \frac{\alpha^n}{\sqrt{n!}} \ket{n}.
\end{equation}
But Eve does not have to be limited by this assumption and theoretically can prepare any state of light. Moreover, infinite-dimensional coherent state complicates the fidelity calculation (especially for imperfect state preparation) needed to determine the phase error that affects the key rate. Therefore, this section provide framework for arbitrary probing state with simplification of fidelity calculation and, accordingly, determining the phase error.

Firstly, let's assume that Eve is able to prepare a Trojan probing with an arbitrary state as presented, for example, in \cite{Pereira22}. But, unlike \cite{Pereira22}, this paper focus on detailed consideration Trojan probing of PM and corresponding usage of fidelity between quantum states. It is important to mention the interpretation of Alice behavior when she randomly choose \textit{firstly} a number of photon in pulse \cite{Lim14,Tupkary24} (and only then randomly choose a label of decoy or signal intensity). This paper introduce similar logic: there is assumed that Eve \textit{firstly} decide to obtain a specific number of probing photons reflected from Alice devices and Eve's choice can be described as probability distribution $P_n$.

Let the eavesdropper emission reflected from Alice's devices be given by an arbitrary pure state $\ket{\psi_{\rm out}}$:
\begin{equation}
    \ket{\psi_{\rm out}} = \sum_{n=0}^{\infty} \sqrt{P_n} \ket{n}.
    \label{eq:trojan}
\end{equation}
Hence, the Trojan light can be described as a random variable, the outcome of which is the emission of $n$ photons $\ket{n}$ with probability $P_n$. The eavesdropper attack is bounded in intensity from above by the value $\mu_{\rm out}$ due to the maximum intensity permissible for the integrity of optical fiber \cite{Lucamarini15, Makarov24} (i.e., up to critical intensity of light in optical fiber and optical devices after which the optical elements will be degradate and any transmission will be impossible). $\ket{\psi_{\rm out}}$ has also 4 possible variations, depending on 4 settings on the transmitter phase modulator, which set the states $\rhobar_i$ \eqref{eq:mixed}.

\textbf{Assumption.} \textit{Let the protocol and its implementation be subject to the condition $\mu_{\rm out}<1$, where $\mu_{\rm out}$ is the maximum intensity of the output Trojan emission.}

The choice of such a limitation is due to the fact that the presence of a non-zero key rate requires low values of the output probing intensity $\mu_{\rm out}$ \cite{Lucamarini15}. 

In the case of the implementation of an arbitrary Trojan emission (for now, it means Trojan pure states with arbitrary photon-number distribution), there are random sets of $n_i$ probing photons and each set is obtained $i$-th times from Alice's setup. The total number of photons is limited by:
\begin{equation}
    \sum_{i = 1}^N n_i = N \mu_{\rm out} \,,
\end{equation}
where $N$ is the number of pulses transmitted by Alice during the quantum key distribution. It is worth noting to say that $N \mu_{\rm out}<N$ follows by Assumption, which means that it is impossible to randomly fill all Alice's pulses with non-zero Trojan photons. Each pulse with a non-zero Trojan photon means a pulse compromised by attack and this affects the final length of the secret key, but zero Trojan photons in the $i$-th pulse do not provide an information leakage. It follows that the more pulses are filled (and compromised by Trojan emission), the more information will be received by an eavesdropper through the Trojan side channel.

Consider the following side channel:
\begin{equation}
    \ket{\psi_{\rm s. ch.}} = \sqrt{1-\mu_{\rm out}} \ket{ \rm vac} + \sqrt{\mu_{\rm out}} \ket{1{\,\rm ph.}}\,,
    \label{eq:s.ch.ideal}
\end{equation}
where $\ket{\rm vac}$ is the vacuum state, $\ket{1{\,\rm ph.}}$ is the state with one photon. Such side channel will have an average intensity equal to $\mu_{\rm out}$. The proposed state corresponds to the situation when the Alice's pulses are filled either with one Trojan photon or remain empty. The total number of filled positions will be $N \mu_{\rm out}$.

\textbf{Statement 1.} \textit{Asymptotically (for $N\rightarrow +\infty$), any Trojan emission with arbitrary representation \eqref{eq:trojan} and with average intensity no higher than $\mu_{\rm out}$ gives the eavesdropper less equal or information than side channel \eqref{eq:s.ch.ideal}.}

\begin{proof}
The amount of information received by the eavesdropper is determined by the Alice's pulses that contained Trojan photons. The more such filled pulses means the more information for the eavesdropper. Let us define the number of filled pulses $\ket{\psi_{\rm out}}$ as $K$. The number of filled pulses $K$ will be:
\begin{equation}
  K = \sum_{n=1}^\infty N P_n \leq N \sum_{n=1}^\infty P_n \cdot n \leq N \sum_{n=0}^\infty P_n \cdot n = N \bra{\psi_{\rm out}}\hat{N}\ket{\psi_{\rm out}} \leq N \mu_{\rm out}\,, 
  \label{eq:proof}
\end{equation}
where $\bra{\psi_{\rm out}}\hat{N}\ket{\psi_{\rm out}}$ is average intensity of $\ket{\psi_{\rm out}}$ and it is taken into account that $N\rightarrow +\infty$. $N \mu_{\rm out}$ is the number of filled pulses for \eqref{eq:s.ch.ideal}, i.e., due to the inequality \eqref{eq:proof}, side channel \eqref{eq:s.ch.ideal} gives more or equal information to the eavesdropper than arbitrary Trojan probing \eqref{eq:trojan}.
\end{proof}

However, in practical implementations of QKD, the statistics of measured observables are finite and the condition $N\rightarrow +\infty$ is not satisfied. In addition, due to channel losses and the usage of weak coherent
states with significant vacuum component, only $M_1^L<N$ pulses are used to generate the secret key (where $M_1^L$ is the decoy-state method estimation of a number of single-photon bits in the verified key). All other pulses do not provide information about the key and their leakage will not lead to the disclosure of bits of the verified key. Accordingly, the substitution $N\rightarrow M_1^L$ will be used in the proof of the following statement.

\textbf{Statement 2.} \textit{With an accuracy of $2\varepsilon$, any Trojan attack on PM of form \eqref{eq:trojan} and with an intensity no higher than $\mu_{\rm out}$ gives the eavesdropper less information than side channel with form:}
\begin{equation}
    \ket{\psi_{\rm s. ch.}} = \sqrt{1-\mu_{\rm out}'} \ket{ \rm vac} + \sqrt{\mu_{\rm out}'} \ket{1{\,\rm ph.}}\,,
    \label{eq:s.ch.}
\end{equation}
\textit{where $\mu_{\rm out}'$ is is determined through the implicit equation $M_1^L \mu_{\rm out}' - \delta^{L}(M_1^L \mu_{\rm out}') = M_1^L \mu_{\rm out} + \delta^{U}(M_1^L \mu_{\rm out})$ and $\delta^{L,U}( x)$ is statistical corrections of the Chernoff bound% или неравенства Азумы
, $M_1^L$ is lower bound for single-photon clicks.}

\textit{Remark.} The upper bound $K_1^U$ of the number of single-photon clicks compromised by a probing attack  is defined similarly \eqref{eq:proof}:
\begin{equation}
  K_1^U =  \mathbb{E} [K_1] + \delta^{U}( K_1) = M_1^L\sum_{n=1}^\infty P_n + \delta^{U}( \mathbb{E} [K_1]) \leq M_1^L \mu_{\rm out} + \delta^{U}(\mathbb{E} [K_1])\,, 
  \label{eq:proof2}
\end{equation}
where used
\begin{equation}
    \sum_{n=1}^\infty P_n \leq \sum_{n=1}^\infty n P_n
\end{equation}
and $\mu_{\rm out} = \sum_{n=1}^\infty n P_n $ is upper bound for the average Trojan output intensity for $\ket{\psi_{\rm out}}$ and $\delta^{U}( x)$ is provided by the Chernoff bound. 
\begin{proof}
It can be shown for the Chernoff bounds that for two expectations $0\leq A \leq B$:
\begin{equation}
    f(A) \leq f(B)\,
\end{equation}
where $f(x) = x+\delta^{U}(x)$, i.e. the inequality of expectations provides the inequality for the value bounds (the proof is given in the Appendix \ref{app:l_sec} in Lemma 1).
Let $B= M_1^L \mu_{\rm out}$ and $A=  \mathbb{E} [K_1] \leq M_1^L \mu_{\rm out}$, then the inequality \eqref{eq:proof2} is rewritten as:
\begin{equation}
  K_1^U \leq M_1^L \mu_{\rm out} + \delta^{U}(M_1^L \mu_{\rm out})\,, 
  \label{eq:proof3}
\end{equation}
The expectation of compromised single-photon clicks for a side channel of the form \eqref{eq:s.ch.} is defined as $\mathbb{E} [K_{s.ch.}] = M_1^L \mu_{\rm out}'$. Similarly to \eqref{eq:proof3}, the lower bound is estimated as:
\begin{equation}
K_{s.ch.}^L \geq M_1^L \mu_{\rm out}' - \delta^{L}(M_1^L \mu_{\rm out}')\,, \label{eq:proof4}
\end{equation}
and let this estimation be greater than the upper estimation of single-photon clicks compromised by the Trojan attack:
\begin{equation}
   K_1^U \leq K_{s.ch.}^L\,.
\end{equation}
This is satisfied by the following condition on $\mu_{\rm out}'$:
\begin{equation}
     M_1^L \mu_{\rm out}' - \delta^{L}(M_1^L \mu_{\rm out}') = M_1^L \mu_{\rm out} + \delta^{U}(M_1^L \mu_{\rm out})\,,
\end{equation}
which proves the original statement with an accuracy of $2\varepsilon$. \end{proof}

Still we assume pure states for Trojan probing. Now we write a generic (possibly mixed) state for Eve's probing system:
\begin{equation}
     \rho_{\rm out} = \sum_{m} p_m \ket{\psi_m} \bra{\psi_m},
     \label{eq:trojan_mixed}
\end{equation}
where $\ket{\psi_m}$ can be not mutually orthogonal and has form close to \eqref{eq:trojan}:
\begin{equation}
    \ket{\psi_m} = \sum_{n=0}^{\infty} \sqrt{P_{n,m}} \ket{n}.
\end{equation}
Any $\rho_{\rm out}$ can be purified:
\begin{equation}
     \ket{\psi_{\rm out}'} = \sum_{m} \sqrt{p_m} \ket{\psi_m} \ket{a_m} = \sum_{m} \sum_{n=0}^{\infty} \sqrt{p_m P_{n,m}} \ket{n} \ket{a_m},
\end{equation}
where $\{\ket{a_m}\}$ is orthonormal basis of Eve's ancillary system. The number of vacuum Trojan pulses can be obtained as:
\begin{equation}
     K_0 = N \sum_{m} p_m P_{0, m} = N - N \sum_{m} p_m \sum_{n=1}^{\infty} P_{N, m}
\end{equation}
and filled Trojan pulses is 
\begin{equation}
     K = N \sum_{m} p_m \sum_{n=1}^{\infty} P_{N, m}
\end{equation}
It is allow us to write equation similar to \eqref{eq:proof}:
\begin{equation}
\begin{split}
  K &= N  \sum_{n=1}^{\infty} \sum_{m} p_m P_{N, m} \leq N \sum_{n=1}^\infty n \cdot \sum_{m} p_m P_{n,m} \\
  &\leq N \sum_{n=0}^\infty n \cdot \sum_{m} p_m P_{n,m} = N \bra{\psi_{\rm out}'}\hat{N}\ket{\psi_{\rm out}'} = {\rm Tr} [\hat{N} \rho_{\rm out} ] \leq N \mu_{\rm out}\,. 
  \label{eq:mixed_K}
\end{split}
\end{equation}
Note, that operator $\hat{N}$ does not act on ancillary system of Eve. With the equation \eqref{eq:mixed_K}, Statement 1 and Statement 2, we can write final generalized statement.

\textbf{Statement 3.} \textit{With an accuracy of $2\varepsilon$, any Trojan attack on PM of arbitrary form \eqref{eq:trojan_mixed} and with an intensity no higher than $\mu_{\rm out}$ gives the eavesdropper less or equal information than side channel with form:}
\begin{equation}
    \ket{\psi_{\rm s. ch.}} = \sqrt{1-\mu_{\rm out}'} \ket{ \rm vac} + \sqrt{\mu_{\rm out}'} \ket{1{\,\rm ph.}}\,,
\end{equation}
\textit{where $\mu_{\rm out}'$ is is determined through the implicit equation $M_1^L \mu_{\rm out}' - \delta^{L}(M_1^L \mu_{\rm out}') = M_1^L \mu_{\rm out} + \delta^{U}(M_1^L \mu_{\rm out})$ and $\delta^{L,U}( x)$ is statistical correction of the Chernoff bound% или неравенства Азумы
, $M_1^L$ is lower bound for single-photon clicks.}

Now we found that any Trojan attack with average intensity ${\rm Tr} [\hat{N} \rho_{\rm out} ] \leq \mu_{\rm out}$ can be conservatively bounded by a side channel of the form \eqref{eq:s.ch.} with $2\varepsilon$-accuracy. The side channel \eqref{eq:s.ch.} can be represented in the case of polarization encoding as four density matrices as:
\begin{equation}
\begin{split}
    \rho_{E,1} &= 
    \begin{pmatrix}  
        \frac{\mu_{\rm out}'}{2}& \frac{\mu_{\rm out}'}{2}& \frac{1}{2}\sqrt{\mu_{\rm out}'(1 - \mu_{\rm out}')}\\
        \frac{\mu_{\rm out}'}{2}& \frac{\mu_{\rm out}'}{2}& \frac{1}{2}\sqrt{\mu_{\rm out}'(1 - \mu_{\rm out}')}\\
        \frac{1}{2}\sqrt{\mu_{\rm out}'(1 - \mu_{\rm out}')}& \frac{1}{2}\sqrt{\mu_{\rm out}'(1 - \mu_{\rm out}')}& 1 - \mu_{\rm out}'\\
    \end{pmatrix} \,,\\
    \rho_{E,2} &= 
    \begin{pmatrix}  
        \frac{\mu_{\rm out}'}{2} &-\frac{\mu_{\rm out}'}{2}&  \frac{1}{2}\sqrt{\mu_{\rm out}'(1 - \mu_{\rm out}')}\\
        -\frac{\mu_{\rm out}'}{2}& \frac{\mu_{\rm out}'}{2}& -\frac{1}{2}\sqrt{\mu_{\rm out}'(1 - \mu_{\rm out}')}\\
        \frac{1}{2}\sqrt{\mu_{\rm out}'(1 - \mu_{\rm out}')}& -\frac{1}{2}\sqrt{\mu_{\rm out}'(1 - \mu_{\rm out}')}& 1 - \mu_{\rm out}'\\
    \end{pmatrix} \,,\\
    \rho_{E,3} &= 
    \begin{pmatrix}  
        \frac{\mu_{\rm out}'}{2} & -\frac{i\mu_{\rm out}'}{2}& \frac{1}{2}\sqrt{\mu_{\rm out}'(1 - \mu_{\rm out}')}\\
        \frac{i\mu_{\rm out}'}{2}& \frac{\mu_{\rm out}'}{2}  & \frac{i}{2}\sqrt{\mu_{\rm out}'(1 - \mu_{\rm out}')}\\
        \frac{1}{2}\sqrt{\mu_{\rm out}'(1 - \mu_{\rm out}')} & -\frac{i}{2}\sqrt{\mu_{\rm out}'(1 - \mu_{\rm out}')}& 1 - \mu_{\rm out}'\\
    \end{pmatrix} \,,\\
    \rho_{E,4} &= 
    \begin{pmatrix}  
        \frac{\mu_{\rm out}'}{2} &\frac{i\mu_{\rm out}'}{2}& \frac{1}{2}\sqrt{\mu_{\rm out}'(1 - \mu_{\rm out}')}\\
        -\frac{i\mu_{\rm out}'}{2}& \frac{\mu_{\rm out}'}{2} & -\frac{i}{2}\sqrt{\mu_{\rm out}'(1 - \mu_{\rm out}')}\\
        \frac{1}{2}\sqrt{\mu_{\rm out}'(1 - \mu_{\rm out}')} & \frac{i}{2}\sqrt{\mu_{\rm out}'(1 - \mu_{\rm out}')}& 1 - \mu_{\rm out}'\\
    \end{pmatrix} \,.
\end{split}
\label{eq:matrix_e}
\end{equation}
where, for example, the first matrix is obtained from $\ket{\psi_{\rm s. ch., 1}} \bra{\psi_{\rm s. ch., 1}}$:
\begin{equation}
    \ket{\psi_{\rm s. ch., 1}} = \sqrt{1-\mu_{\rm out}'} \ket{ \rm vac} + \sqrt{\mu_{\rm out}'} \ket{D_{1\,\rm ph.}}
    \label{eq:psi_s_ch_1}
\end{equation}
and $\ket{D_{1\,\rm ph.}} = (\ket{H_{1\,\rm ph.}} + \ket{V_{1\,\rm ph.}})/\sqrt{2}$ denotes diagonal polarization with one photon. $\ket{\psi_{\rm s. ch., 1}} $ limits from above the Trojan light that came out of the Alice's setup when the Alice's pulse $\rhobar_1$ is transmitted. All density matrices are found similarly to \eqref{eq:psi_s_ch_1}.

The density matrices \eqref{eq:matrix_e} correspond %(up to a phase shift $\varphi_0$) 
to the side channel emission when the phase modulator is configured to generate states $\ket{\psi_1},\, \ket{\psi_2},\, \ket {\psi_3}$ and $\ket{\psi_4}$, respectively. Then the light in different polarization bases emitted from Alice's setup can be conservatively estimated from above as $\rho_{BE, \Xp}$ and $\rho_{BE, \Yp}$:
\begin{equation}
    \begin{split}
        \rho_{BE, \Xp} &= \rhobar_1 \otimes \rho_{E,1} + \rhobar_2 \otimes \rho_{E,2}\, , \\
        \rho_{BE, \Yp} &= \rhobar_3 \otimes \rho_{E,3} + \rhobar_4 \otimes \rho_{E,4}\, .
    \end{split}
\end{equation}
(Here a generic form of the Alice's states $ \rhobar_i$ is choosed due to possible imperfect state preparation described in Appendices \ref{app:imp_prep} and \ref{app:gauss}.)

To quantify the difference between the phase error $E_1^{\ph,\Xp}$ and the bit error $E_1^\Yp$, let use the concept of quantum coin introduced in~\cite{GLLP} for an equivalent entanglement-based virtual protocol. Applying the complementarity argument \cite{Koashi09} and the Bloch sphere bound \cite{Tamaki03} to the quantum coin yields the following inequality \cite{Lo07}:
\begin{equation}
    \sqrt{F(\rho_{BE, \Xp}, \rho_{BE, \Yp})} \leq 1 - \Y_1 + \Y_1 \bigg(\sqrt{E_1^{\ph,\Xp} E_1^\Yp} + \sqrt{(1 - E_1^{\ph,\Xp})(1 - E_1^\Yp)}\bigg) \,,
    \label{eq:F_inequality}
\end{equation} 
where the conditional probability of single-photon clicks $\Y_1=(\Y_1^\Xp+\Y_1^\Yp)/2$ is introduced and the fidelity $F$ between two states is defined as:
\begin{equation}
	F(\rho_{BE, \Xp}, \rho_{BE, \Yp}) \equiv \bigg[\text{Tr}\bigg(\sqrt{\sqrt{\rho_{BE, \Xp}} \, \rho_{BE, \Yp} \, \sqrt{\rho_{BE, \Xp}}}\bigg)\bigg]^2 \,.
    \label{eq:F_alt}
\end{equation}
Here we have not taken into account possible transformations of all matrices by the same unitary operator $U$ to Eve's and Alice's system, but it can be shown (through properties of square root of matrices and trace and properties of self-adjoint and unitary matrices) that substituting $U$ into \eqref{eq:F_alt} will not change the value of $F(\rho_{BE, \Xp}, \rho_{BE, \Yp})$.
Solving \eqref{eq:F_inequality}, one can obtain the following upper bound on the single-photon phase errors:
 \begin{equation}
    E_1^{\ph,\Xp} \leq E_1^\Yp + 4 \Delta^\prime (1 - \Delta^\prime) (1 - 2E_1^\Yp) + 4(1 - 2\Delta^\prime) \sqrt{\Delta^\prime (1 - \Delta^\prime) E_1^\Yp (1 - E_1^\Yp)} \,,
    \label{eq:E1_coin_alt}
\end{equation}
\begin{equation}
    \Delta^\prime = \frac{1 - \sqrt{F(\rho_{BE, \Xp}, \rho_{BE, \Yp})}}{2\Y_1} = \frac{\Delta}{\Y_1} \,,
\end{equation}
where the value $\Delta=(1-\sqrt{F})/2$ is usually called the quantum coin imbalance.

% \red{Maybe add the $\ket{\Psi}_{ABE}$ like Lucamarini and Curty and their phase error estimation using the scalar product of purifications since there are a comparison with the previous THA methods.}

\begin{figure}[t!]\centering
\includegraphics[width=0.49\textwidth]{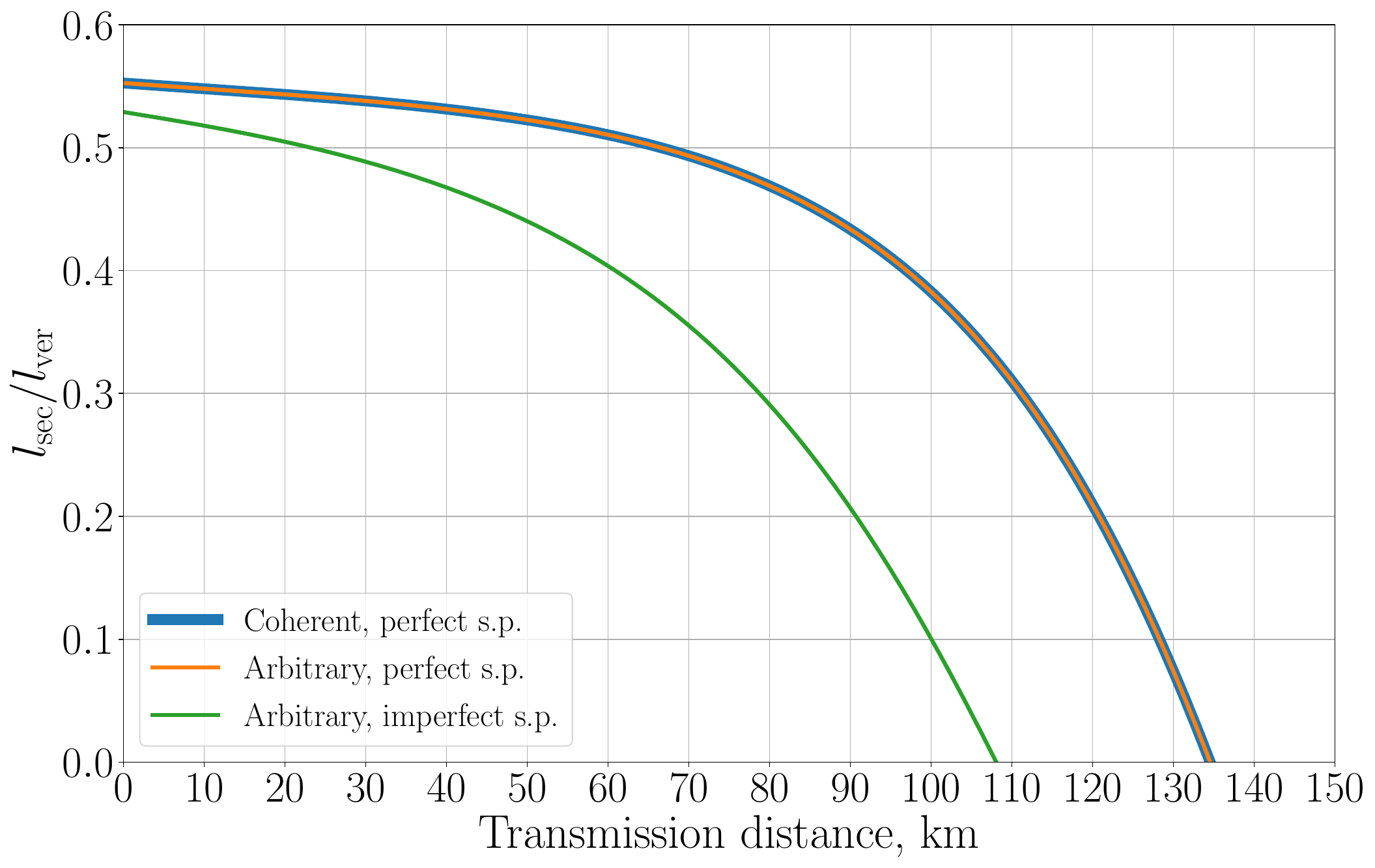}
    \includegraphics[width=0.49\textwidth]{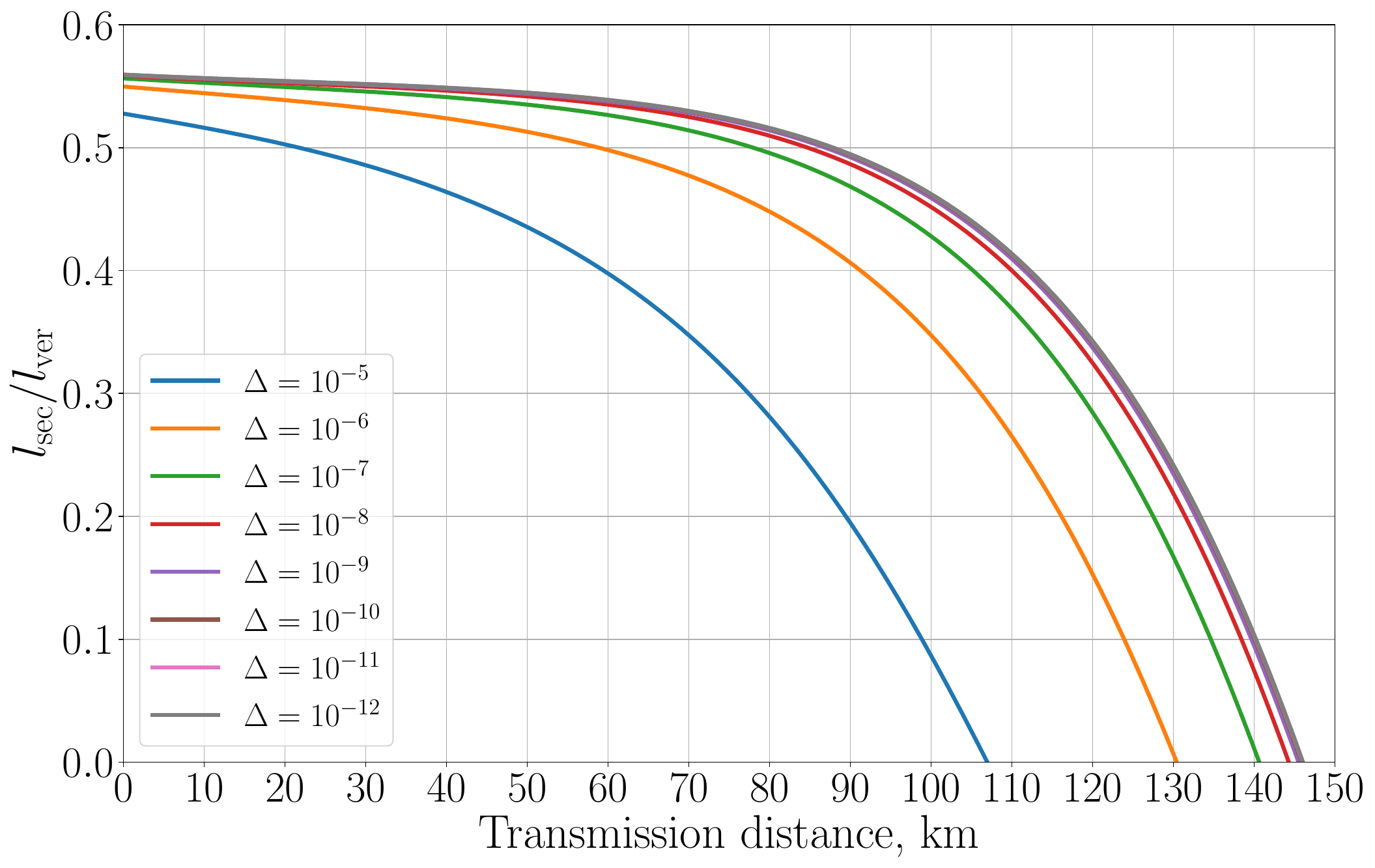}
	\caption{Left: the secret key rate per one verified click. Blue line indicates the scenario for the Trojan attack by probing with coherent pulses in the presence of perfect Alice's state preparation ($\Delta = 5 \times 10^{-7}$), orange and green are for Trojan probing by arbitrary pure states with perfect ($\Delta = 5 \times 10^{-7}$) and imperfect ($\Delta=9.2 \times 10^{-6}$) Alice's state preparations respectively. Right: the secret key rate per one verified click for different values of $D\Delta$. Note, configurations of three isolators $\{28 { \rm \,dB},\,28 { \rm \,dB},\,48 { \rm \,dB}\}$ and $\{28 { \rm \,dB},\,48 { \rm \,dB},\,48 { \rm \,dB}\}$ provide (for the case of ideal state preparation) $\Delta \approx 10^{-9}$ and $\Delta \approx 10^{-11}$ respectively.}
	\label{fig:r_sec}
\end{figure}

\section{Results and discussion}
The maximum input intensity of the Trojan probing is chosen as $I_{\rm in} = 2.5 \times 10^{12}$ photons per Alice's pulse and this value was determined for commercial QKD scheme \cite{Makarov24}. The total loss of the Trojan photons in the Alice's equipment is $\alpha_{\rm A} = 172 { \rm \,dB}$, where two optical isolators with $28 { \rm \,dB}$ and $48 { \rm \,dB}$ is used as passive countermeasure against Trojan-horse attack (see \cite{Makarov24} for detailed description of the commercial scheme and experimental analysis of its Trojan-horse loophole). Authors determine the Trojan output intensity as $\mu_{\rm out} = 10^{-\alpha_{\rm A}/10} I_{\rm in} = 1.5 \times 10^{-5}$. 
In  this paper %$\mu_{\rm out} = 0$ and also 
$\mu_{\rm out}= 10^{-6}$ is chosen, which is realistic value for scheme with 2-3 isolators and corresponds previous investigations \cite{Lucamarini15, Pereira19}. However, the work \cite{Makarov24} suggest to use an additional isolator for achieving value $\mu_{\rm out} \leq 10^{-9}$. Therefore, the values up to $\mu_{\rm out}= 10^{-12}$ are considered in the context of an influence on the secrete key rate.

Two cases of Alice's states are examined: the ideal state preparation \eqref{eq:rho_ideal} and the imperfect one with the Gaussian approximation \eqref{eq:rho_i_averaged}. The imbalance value of the quantum coin were found as $\Delta = 5 \times 10^{-7}$ and $\Delta=9.2 \times 10^{-6}$  by new approach and with the states \eqref{eq:rho_ideal} and \eqref{eq:rho_i_averaged}, respectively. $\Delta=5 \times 10^{-7}$ is for Trojan attack by coherent states \cite{Lucamarini15} and perfectly prepared states. Simulation of the secret key rates gives the results shown in Fig.~\ref{fig:r_sec} (see Subsection 2.1 Experimental Setup and Appendix B in \cite{Reutov23} for a description of the simulation methods and parameters).

It is worth noting to say that there are presented a rather conservative approach by replacing one side channel (Trojan attack with emission given by arbitrary statistics) with another side channel  given by \eqref{eq:s.ch.}. However, the evaluation for the conservative side-channel \eqref{eq:s.ch.} yields almost the same secrete key rate as in the scenario with the additional assumption of coherent Trojan states. The secrete key rate is also estimated for realistic setup parameters and imperfect state preparation (for simplicity described by a Gaussian model of imperfect polarization-state preparation) and the key transmission maximum range is found higher than 100 km (see left Fig.~\ref{fig:r_sec}).

The main contribution to the quantum-coin imbalance value is derived from imperfect state preparation and it can be seen in significant difference between $\Delta$'s for ideal state preparation and modeled imperfect one. For example, the value $\mu_{\rm out} = 10^{-100}$ provide $\Delta = 8.8 \times 10^{-6}$ for the proposed model of imperfectly prepared states. Nevertheless, after  $\Delta < 10^{-8}$ (right Fig.~\ref{fig:r_sec}) the influence on the key rate decrease rapidly. Consequently, the main leakage threat comes from imperfect preparation and conservative estimation of the Trojan-horse attack by side channel \eqref{eq:s.ch.} (instead of previous approaches \cite{Lucamarini15, Wang18, Tamaki16}) does not significantly reduce the secrete key rate.

\section{Conclusions}\label{sec:conclusions}
This work provides a  analysis of the security of quantum key distribution (QKD) systems against Trojan horse attack on phase modulators, considering both ideal and imperfect state preparation. By addressing imperfections in transmitted states and modeling Trojan probing by arbitrary generic (pure or mixed) states, a framework for enhancing QKD security in practical scenarios is developed. The Trojan attack model is generalized to arbitrary states, conservative bounds on information leakage are derived as leakage through single-photon side-channel states. This approach allows to more realistically assess the threat posed by Trojan horse attacks and it does not rely on restrictive assumptions about the nature of the eavesdropper's probing pulses. 

The quantum coin imbalance remains low ($5\times 10^{-7}$ for ideal and $9.2\times10^{-6}$ for imperfect preparation) even under Trojan attack on PM, demonstrating the resilience of practical QKD systems when equipped with proper countermeasures, such as isolators and spectral filters, which are essential for mitigating the risk of information leakage. Moreover, numerical simulations confirm secure key distribution over distances higher than 100 km. This work bridges theory and practice, offering tools to enhance QKD security against emerging threats for practical applications and increasing the feasibility of quantum-secured communication. 
%In conclusion, this study advances the understanding of QKD vulnerabilities and provides a foundation for designing secure, real-world quantum communication systems.

\section*{Acknowledgements}
The author is grateful to Andrey Tayduganov and Daniil Menskoy for fruitful discussions and valuable advices on the manuscript.

\begin{appendices}

\section{Imperfect state preparation}\label{app:imp_prep}

Density matrices $\rho_\Xp=\frac{1}{2}(\ketbra{\psi_1}{\psi_1 }+\ketbra{\psi_2}{\psi_2})$ and $\rho_\Yp=\frac{1}{2}(\ketbra{\psi_3}{\psi_3}+\ketbra{\psi_4}{\psi_4 })$ can be written as follows in the case of imperfect preparation:
\begin{equation}
    \begin{split}
        \rho_\Xp &= \frac{1}{2}
        \begin{pmatrix}
            1+\cos\theta & e^{-i\frac{\varphi_1+\varphi_2}{2}} \cos\big(\frac{\varphi_1-\varphi_2}{2}\big) \sin\theta \\
            e^{i\frac{\varphi_1+\varphi_2}{2}} \cos\big(\frac{\varphi_1-\varphi_2}{2}\big) \sin\theta & 1-\cos\theta
        \end{pmatrix} \,, \\
        \rho_\Yp &= \frac{1}{2}
        \begin{pmatrix}
            1+\cos\theta & e^{-i\frac{\varphi_3+\varphi_4}{2}} \cos\big(\frac{\varphi_3-\varphi_4}{2}\big) \sin\theta \\
            e^{i\frac{\varphi_3+\varphi_4}{2}} \cos\big(\frac{\varphi_3-\varphi_4}{2}\big) \sin\theta & 1-\cos\theta
        \end{pmatrix} \,.
    \end{split}
    \label{eq:rho_XY}
\end{equation}
The condition \eqref{eq:psi_ideal} $\rho_\Xp=\rho_\Yp$ is satisfied for perfectly prepared states, i.e. the photon source is basis-independent. If $\rho_\Xp\neq\rho_\Yp$ is true, then the values of single-photon phase errors $E_1^{\ph,\Xp}$ in $\Xp$-basis are not estimated from the measured bit error $E_1^\Yp$ in $\Yp$-basis (or vice versa).

In the paper ~\cite{Wang23} devoted to fully passive QKD, the authors consider an equivalent protocol based on virtual entanglement. In this protocol, the source emits signal states with mixed polarization in the $Z$-basis. It is argued that Alice's imperfect preparation of states in the $Z$-basis is equivalent to Bob's imperfect measurement. This leads to one of the ideas in ~\cite{Wang23, Zapatero23} -- the replacing of the source of randomly fluctuating on Bloch sphere pure states $\{\ket{\psi_i}\}$ with an equivalent source emitting mixed states $\{\rhobar_i\}$,
\begin{equation}
    \rhobar_i = \int_0^{2\pi} \int_0^\pi p_i(\varphi, \theta) \ketbra{\psi(\varphi, \theta)}{\psi(\varphi, \theta)} d\varphi d\theta \,,
    \label{eq:mixed}
\end{equation}
where distributions $\{p_i(\varphi,\theta)\}$ is characterized directly from the test experimental setup measurements (a simple Gaussian approximation of this distributions is proposed in Appendix \ref{app:gauss}). The density matrices \eqref{eq:rho_XY} in this case are replaced by another one:
\begin{equation}
    \rhobar_\Xp = \rhobar_1 + \rhobar_2 \,, \quad
    \rhobar_\Yp = \rhobar_3 + \rhobar_4 \,.
    \label{eq:rho_XY_averaged}
\end{equation}

\section{Gaussian distribution $p_i(\varphi, \theta)$}\label{app:gauss}
Using a simple Gaussian model $p_i(\varphi, \theta) = G\big(\varphi, \phibar_i, \sigma_{\varphi_i}\big) G\big(\theta, \thetabar, \sigma_\theta\big)$, the following analytical approximation can be obtained:
\begin{equation}
    \begin{split}
        \rhobar_i &\simeq \int_0^{2\pi} \int_0^\pi G\big(\varphi, \phibar_i, \sigma_{\varphi_i}\big) G\big(\theta, \thetabar, \sigma_\theta\big) \ketbra{\psi(\varphi, \theta)}{\psi(\varphi, \theta)} d\varphi d\theta \\
        &\simeq \frac{1}{2}
        \begin{pmatrix}
        1 + e^{-\frac{\sigma_\theta^2}{2}} \cos\thetabar
        & e^{-i\phibar_i - \frac{1}{2} (\sigma_{\varphi_i}^2 + \sigma_\theta^2)} \sin\thetabar \\ 
        e^{i\phibar_i - \frac{1}{2} (\sigma_{\varphi_i}^2 + \sigma_\theta^2)} \sin\thetabar
        & 1 - e^{-\frac{\sigma_\theta^2}{2}} \cos\thetabar
        \end{pmatrix} \,,
    \end{split}
	\label{eq:rho_i_averaged}
\end{equation}
where for simplicity the probability density normalization coefficients are omited, since
\begin{equation}
    \int_0^{x_{\max}} G(x,\overline{x},\sigma_x)dx = \frac{1}{2}\bigg[\erf\bigg(\frac{\overline{x}}{\sqrt2\sigma}\bigg) + \erf\bigg(\frac{x_{\max} - \overline{x}}{\sqrt2\sigma}\bigg)\bigg] \simeq 1 \,,
\end{equation}
for $\sigma_x\ll\overline{x}$ and $\sigma_x\ll x_{\max}-\overline{x}$.
% Более точный анализ будет подразумевать использование эмпирических распределений (см. Рис.~\ref{fig:Stokes} и Рис.~\ref{fig:angle_histos}, примеры измерений распределения углов, описывающих поляризацию четырех типов состояний протокола).

\section{Non-decreasing of Chernoff bound}\label{app:l_sec}

\textbf{Lemma 1.} \textit{The following holds:}
\begin{equation}
    f(x)>f(y),
\end{equation}
\textit{for the Chernoff bound $f(x) = (1+\delta)x$}
\begin{equation}
	\bigg[ \frac{e^{\delta}}{(1 + \delta)^{1 + \delta}} \bigg]^x = \varepsilon \,, \quad
	\delta > 0,\, x > 0, \label{eq:chernoff}
\end{equation}
\textit{and for $x>y$. The same statement holds for the lower Chernoff bound.}
\begin{proof}
    Let introduce the notation $f(x) = (1+\delta)x = zx$ and rewrite \eqref{eq:chernoff}:
    \begin{equation}
        \frac{e^{z-1}}{z^z} = \varepsilon^{1/x}\label{eq:lemma_1}
    \end{equation}
    Derivative with respect to $x$ is:
    \begin{equation}
         z^\prime \frac{e^{z-1}}{z^z} {\rm ln}{z} = \frac{\varepsilon^{1/x}}{x^2} {\rm ln}(\varepsilon)
         \label{eq:lemma_11}
    \end{equation}
    Substitute the left side of the equation \eqref{eq:lemma_1} into the right side of \eqref{eq:lemma_11}:
    \begin{equation}
        z^\prime \frac{e^{z-1}}{z^z} {\rm ln}{z} = \frac{e^{z-1}}{z^z} \frac{1}{x^2} {\rm ln}(\varepsilon)
    \end{equation}
        \begin{equation}
        z^\prime {\rm ln}{z} = \frac{1}{x^2} {\rm ln}(\varepsilon) \label{eq:lemma_2}
    \end{equation}
    Find ${\rm ln}(\varepsilon)$ using \eqref{eq:lemma_1}:
    \begin{equation}
        {\rm ln}(\varepsilon)= x(z-1-z {\rm ln}(z))
    \end{equation}
    and substitute in \eqref{eq:lemma_2}:
    \begin{equation}
        z^\prime {\rm ln}{z} = \frac{z-1-z {\rm ln}(z)}{x} 
    \end{equation}
        \begin{equation}
        z^\prime = \frac{z-1-z {\rm ln}(z)}{x{\rm ln}{z}} 
    \end{equation}
    Derivative $f^\prime(x)$ is:
    \begin{equation}
        f^\prime(x)= xz^\prime(x) + z(x) = \frac{z-1-z {\rm ln}(z)}{{\rm ln}{z}}+z = \frac{z-1}{{\rm ln}{z}}
    \end{equation}
    Since $\delta>0$, then $z=1+\delta>1$ and ${\rm ln(z)}>0$ which means $f^\prime(x)= (z-1)/{\rm ln}{z}>0$. Consequently, $f(x)$ is an increasing function. Similarly, reversed inequality can be proven for the lower Chernoff bound.
\end{proof}
\textit{Remark.} Also the differential equation \eqref{eq:lemma_2} gives an exact solution for $\delta(x)>0$:
\begin{equation}
    \delta(x)= e^{1+W_0(-\frac{x+{\rm ln}(\varepsilon)}{e x})} - 1 ,
\end{equation}
where $W_0(y)$ is one of the Lambert $W$-function branches. Note, that other Chernoff bounds can be similarly rewritten in an explicit form with Lambert $W$-function branches.
\end{appendices}

\bibliographystyle{utphys}
\bibliography{bibliography}

\end{document}